\documentclass[aps,pra,twocolumn]{revtex4}

\usepackage{amsmath,amssymb}
\def\gfxon{\usepackage[final]{graphicx}}

\gfxon

\usepackage[T1]{fontenc}
\usepackage{amsthm}
\usepackage{amsfonts}
\usepackage{bm}
\usepackage{cancel}
\usepackage[center]{subfigure}

\usepackage{booktabs}
\usepackage{xcolor}

\usepackage{soul}

\makeatletter
\let\old@startsection=\@startsection
\renewcommand{\@startsection}[6]{\old@startsection{#1}{#2}{#3}{#4}{#5}{#6\mathversion{bold}}}
\makeatother

\makeatletter
\def\mr@ignsp#1 {\ifx\:#1\@empty\else #1\expandafter\mr@ignsp\fi}%
\newcommand{\multiref}[1]{\begingroup
\xdef\mr@no@sparg{\expandafter\mr@ignsp#1 \: }%
\def\mr@comma{}%
\@for\mr@refs:=\mr@no@sparg\do{\mr@comma\def\mr@comma{,}\ref{\mr@refs}}%
\endgroup}
\makeatother

\DeclareGraphicsExtensions{.pdf,.png,.jpg}

\ifx\href\asklfhas\newcommand{\href}[2]{#2}\fi

\begin{document}

\title{Vortex lattices in 
three-component Bose-Einstein condensates 
under rotation:
simulating colorful vortex lattices in a color superconductor}

\author{Mattia Cipriani$^{1}$, Muneto Nitta$^{2}$}

\affiliation{$^1$University of Pisa, Department of Physics ``E. Fermi'', INFN,
Largo Bruno Pontecorvo 7, 56127, Italy \\
$^2$Department of Physics, and Research and Education Center 
for Natural Sciences, Keio University, 4-1-1 Hiyoshi, Yokohama, 
Kanagawa 223-8521, Japan
}

\begin{abstract}
We study  vortex lattices
in three-component BECs under rotation,
where three kinds of fractional vortices winding one of three components are present. 
Unlike the cases of two-component BECs 
where the phases of square and triangular lattices 
are present depending on the intercomponent 
coupling constant and the rotation speed, 
we find triangular ordered ``colorful'' vortex
lattices 
where 
three kinds of fractional vortices are placed in 
order without defects, 
in all parameter regions where 
the inter-component coupling $g'$ is less 
than the intra-component coupling $g$.
When $g>g'$ on the other hand, 
we find the phase separation: 
In a region where one component is present, 
the other two components must vanish, 
where we find ghost vortices in these two components 
whose positions are separated. 
In the boundary $g=g'$, the accidental $U(3)$ symmetry is present, in which case two vortices in different components 
are close to each other in some regions.

\end{abstract}

{\footnotesize \texttt{IFUP-TH/2013-11} }

\maketitle

\sloppy

\section{Introduction}
One of the growing topics in condensed matter physics, 
high energy physics, and astrophysics are multi-component condensations.
In this sense, interesting fields of research are
exotic superconductors, multi-component or spinor Bose-Einstein condensates (BECs) 
of ultra--cold atomic gases, superfluid $^3$He,
exciton-polariton condensates, nonlinear optics, 
hadronic matter composed of neutron and proton Cooper pairs, 
and quark matter composed of di-quark condensates 
consisting of quark Cooper pairs.  
In these systems, fractional vortices can be created by rotating superfluids or BECs or by applying magnetic field on superconductors.  
Fractional vortices are characterized by rational or fractional quantized circulations for superfluids or BECs 
and fluxes for superconductors, 
as found in various systems: superfluid $^3$He \cite{Salomaa:1985,Volovik:2003}, 
$p$-wave superconductors \cite{Salomaa:1985,Ivanov:2001,Chun:2007,Jang:2011},   
multigap superconductors  \cite{Tanaka:2001,Babaev:2002,Gurevich:2003,Goryo:2007},
spinor BECs \cite{Ho:1998,Semenoff:2006vv}, 
multi--component BECs
\cite{Mueller:2002,Son:2001td,Kasamatsu:2003,Kasamatsu:2005,
Kasamatsu:2004,Woo:2008,Kasamatsu:2009,
Eto:2011wp,Aftalion:2012,Kuopanportti:2012,Eto:2012rc,
Cipriani:2013nya,Eto:2013}, 
exciton-polariton condensates \cite{exciton,exciton-lattice}, 
nonlinear optics \cite{optics},
and color superconductors as quark matter 
\cite{Balachandran:2005ev,Nakano:2007dr,Eto:2009kg,
Eto:2009bh,Hirono:2012ki,Vinci:2012mc,Cipriani:2012hr}. 

Theoretical and experimental investigations of the properties of vortices are easily done in 
BECs of ultra--cold atomic gases \cite{Pethick:2008}, 
which reveal to be the ideal physical setting among the others for this purpose. 
In the theory side, BECs can be quantitatively well described in the mean-field theory 
by using the Gross-Pitaevskii (GP) equation. 
Experimentally,  
BECs are quite flexible and controllable systems 
being the atomic interaction tunable through 
a Feshbach resonance \cite{Chin:2010}. 
In addition, the condensates can be visualized directly 
by means of optical techniques. 
Thus far, two-component BECs have been realized 
by using the mixture
of atoms with two hyperfine states of 
$^{87}$Rb \cite{Myatt:1997}
or the mixture of two different species of atoms 
\cite{Modugno:2002,Papp:2008,McCarron:2011}.  
In Refs.~\cite{Mueller:2002,Kasamatsu:2003,Kasamatsu:2005,Woo:2008,
Aftalion:2012,Kuopanportti:2012,Cipriani:2013nya}, 
the phase diagram of the vortex lattice forming in 
two-component BECs was studied, 
and a rich variety of lattices was found. 
When the inter-component coupling $g'$ is increased, 
different configurations from an Abrikosov's triangular lattice 
of fractional vortices to 
a square lattice of fractional vortices have been obtained  \cite{Mueller:2002,Kasamatsu:2003,Kasamatsu:2005} 
when $g'<g$. 

On the other hand,
particularly interesting cases in 
high energy physics and astrophysics include 
quark matter  composed of di-quark condensates 
consisting of quark Cooper pairs \cite{Alford:1998mk},  
which may be realized in the core of neutron stars. 
When up, down, strange quarks 
participate in forming Cooper pairs, 
the order parameter $\Delta$ becomes a $3 \times 3$ matrix, 
on which color $SU(3)$ and flavor $SU(3)$ symmetries act 
from left and right, respectively. 
The Ginzburg-Landau description is available in the 
perturbation theory of quantum chromo dynamics.
In the ground state, for which the condensate can be considered to be 
in a diagonal form in color and flavor 
without loss of generality, 
the color and flavor symmetries are spontaneously 
broken down into a diagonal combination of them. 
For this reason, this phase is called the color-flavor locked phase. 
Since color symmetry is spontaneously broken and  
completely screened, 
quark matter in this phase is a color superconductor.
At the same time, it is a superfluid 
since the baryon $U(1)$ symmetry is spontaneously broken. 
Consequently, 
vortices appearing in this phase 
carry both the quantized circulations  
and quantized color magnetic fluxes \cite{Balachandran:2005ev,Nakano:2007dr,Eto:2009kg,Eto:2009bh}.
They are referred to as non-Abelian since the color fluxes 
they carry are non-Abelian magnetic fields. 
The circulation which they carry is quantized to 
be 1/3 of the unit circulation,  
due to three components having non-zero value 
in the ground state.  
The interaction between non-Abelian vortices 
was calculated in \cite{Nakano:2007dr} 
 to be 1/3 of that between integer quantized superfluid vortices, 
which implies the formation of a colorful vortex lattice when 
the color superconductor is rotating 
as in conventional superfluids. 
The phenomenon for which the vortex lattice 
behaves as a polarizer when light passes through it 
\cite{Hirono:2012ki} is one of the interesting consequences 
of the aforementioned properties of quark matter in the color-flavor locked phase. 
Another interesting consequence is 
the existence of a colorful boojum \cite{Cipriani:2012hr}  
at the interface between a color superconductor 
and the confining phase in which all color degrees 
of freedom must be canceled out.
In the latter, 
three color magnetic fluxes, say red, blue, and green, 
with 1/3 circulations are combined together to   
a purely superfluid colorless vortex with the unit circulation, 
which can penetrate into 
the confining phase.
Even though the colorful vortex lattice is one of the most 
important signals of whether the color-flavor locked phase is realized or not
in the core of neutron stars rotating very rapidly, 
there have been no quantitative studies of 
colorful vortex lattices thus far, 
because of large degrees of freedom of 
non-Abelian $SU(3)$ gauge fields. 

The purpose of  this paper is to propose simulating 
certain aspects 
of colorful vortex lattices in a rotating color superconductor 
by using a BEC of ultra--cold atomic gases.
We create 1/3 quantized vortices 
with three different ``colors'' 
by preparing a rotating three-component BEC.
Vortex lattices in three-components BEC have been investigated
in the context of F=1 spinor BEC \cite{Mizushima};
we propose a non--spinor system for this analysis.
Because our purpose is to simulate a color superconductor,
we consider the symmetric case where 
all three intra--component (inter--component) couplings are the same, 
$g_{11}=g_{22}=g_{33} \equiv g$ ($g_{12}=g_{23}=g_{31} \equiv g^\prime$), and all chemical potentials and
masses are the same 
as the case of  the mixture 
of atoms with different hyperfine states. 
Combining a set of three different vortices together 
results in an integer vortex without a ``color.''
These features are the same with 1/3 quantized vortices 
in the color superconductor.
We solve the GP equation for 
a rotating three-component BEC subject to a harmonic trapping potential  
by an imaginary time propagation, 
and find ``ordered'' Abrikosov's triangular lattices 
in all parameter ranges of $g'<g$. 
Here, the term ``ordered'' implies that 
each of the three different fractional vortices winding around 
the three different components 
constitutes an Abrikosov's triangular lattice 
without defects, 
and that the total configuration is also an Abrikosov's triangular lattice.
In other words, we did not find any defects or displacements of 
vortices with different colors. 
This situation is in contrast to the cases of two-component BECs \cite{Mueller:2002,Kasamatsu:2003,Kasamatsu:2005,Woo:2008,
Aftalion:2012,Kuopanportti:2012,Cipriani:2013nya} 
in which there exist an Abrikosov's triangular lattice and  
a square lattice of fractional vortices in $g'<g$, 
depending on the inter-component coupling $g'$ 
as denoted above.
When $g^{\prime}=g$ the zeros of the density of the different components can be overlapped.
However, the comparison between plots of the phase of the order parameter of each component shows that the vortices are not coincident.
In the parameter range $g<g'$ 
of the phase separation, 
we find that vortex sheets are present like the case of two component BECs \cite{Kasamatsu:2009}. 
In a region where one component is present, 
the other two components are absent because of 
the phase separation.
In that region, ghost vortices of these two components are present and they are separated, even if there is no symmetric ordering as found for $g'<g$. 

\def\figwidth{8cm}

\begin{figure*}
\centering
\begin{tabular}{c@{\hskip 0.5cm}c}
	(a) $\delta=0.5$ & (b) $\delta=0.9$ \\
\includegraphics[height=\figwidth]{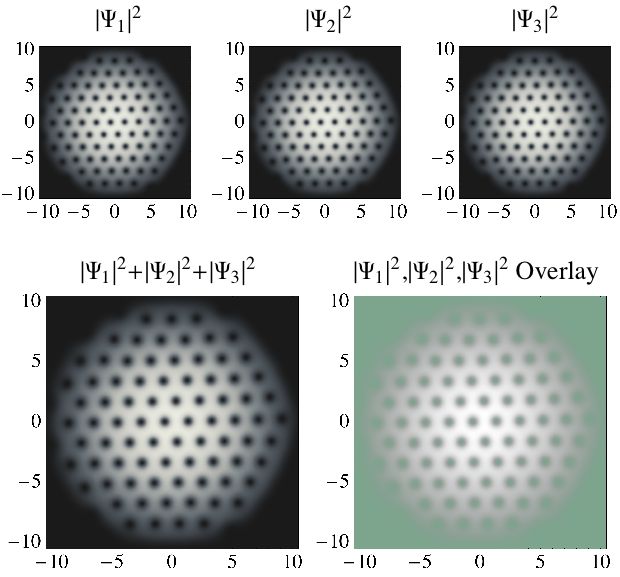} & \includegraphics[height=\figwidth]{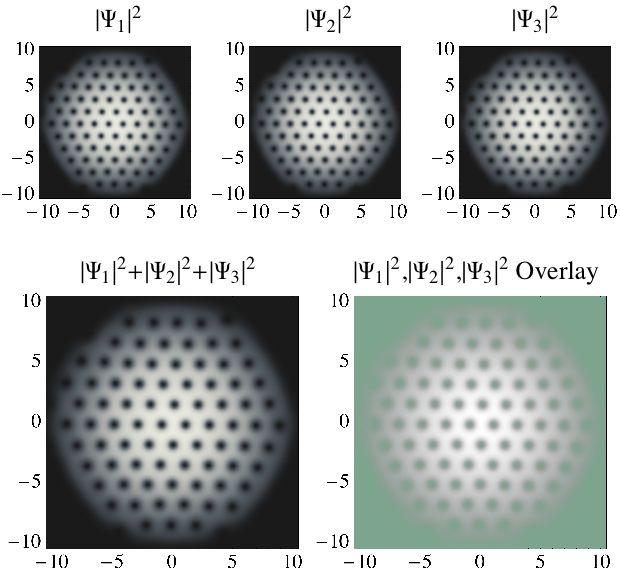} \\ 
	(c) $\delta=1$ & (d) $\delta=1.5$ \\
\includegraphics[height=\figwidth]{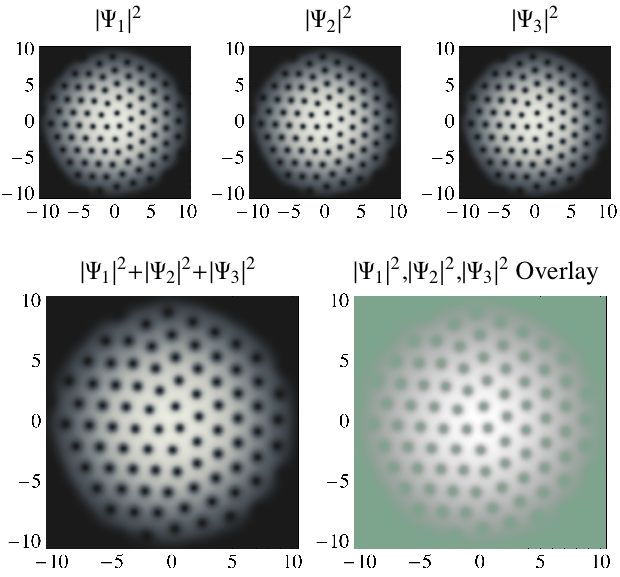} & \includegraphics[height=\figwidth]{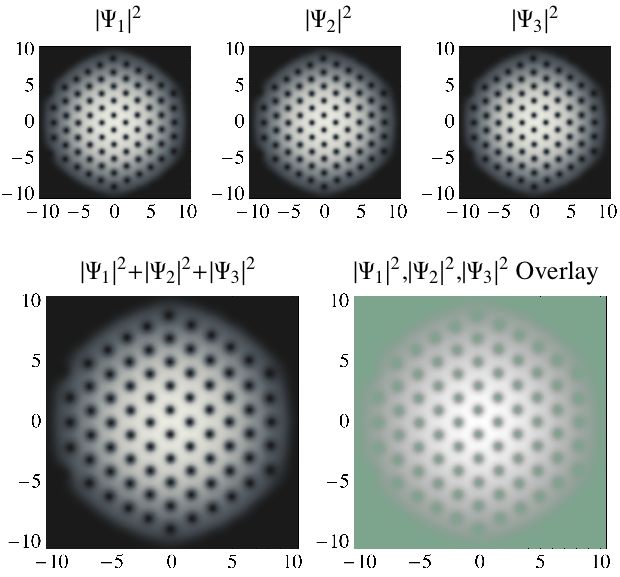} 
\end{tabular}
\caption{
(Color online) 
Some of the results obtained when the initial configuration used in the calculation is the same as the ground state of a non-rotating BEC.
The top pictures in each panel are plots of the density profile for each component, $n_{i}=|\Psi_{i}|^{2}$.
The bottom left picture is the full density profile $n=|\Psi_{1}|^{2} + |\Psi_{2}|^{2} + |\Psi_{3}|^{2}$, while the bottom right picture is the same plot, where the various components are distinguished by color (gray tones).
Red regions correspond to zeros in the density of the first component, 
blue regions stay where the density of the second component vanishes, 
while green areas are the zeros of the third component density.
However, no different colors (gray tones) are visible in this picture because vortices of all three components are coincident. 
The value of $\delta$ is reported on top of each panel.
In all cases we obtain an integer vortex lattice. 
}
\label{fig:Fig1}
\end{figure*}
\def\figwidth{8cm}

\begin{figure*}
\centering
\begin{tabular}{c@{\hskip 0.5cm}c}
	(a) $\delta=-0.1$ & (b) $\delta=-0.3$ \\
\includegraphics[height=\figwidth]{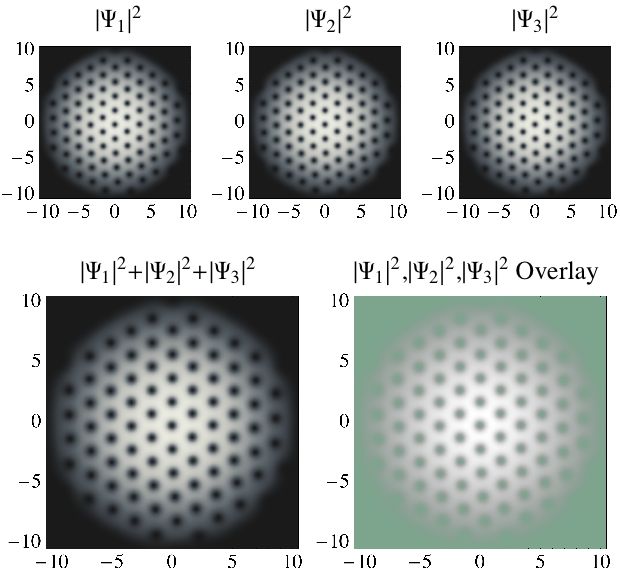} & \includegraphics[height=\figwidth]{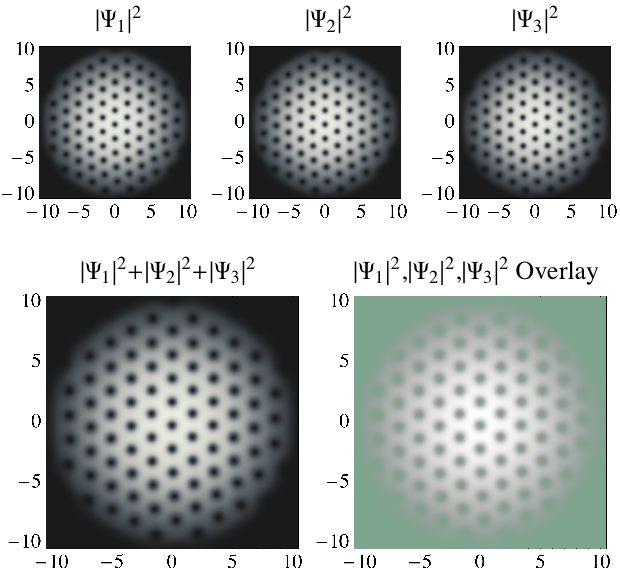} 
\end{tabular}
\caption{
(Color online) 
Integer vortex lattice obtained when $\delta<0$. 
The top pictures in each panel are plots of the single component density $n_{i}=|\Psi_{i}|^{2}$.
The bottom left picture is the full density profile $n=|\Psi_{1}|^{2} + |\Psi_{2}|^{2} + |\Psi_{3}|^{2}$, while the bottom right picture is the same plot, where the various components are distinguished by color (gray tones).
Red regions correspond to zeros in the density of the first component, 
blue regions stay where the density of the second component vanish, 
while green areas are the zeros of the third component density.
However, no different colors (gray tones) are visible in this picture because vortices of all three components are coincident. 
Because of the attracting interaction, integer vortices do not split into fractional vortices. 
}
\label{fig:Fig2}
\end{figure*}
\def\figwidth{8cm}

\begin{figure*}
\centering
\begin{tabular}{c@{\hskip 0.5cm}c}
	(a) $\delta=0.2$ & (b) $\delta=0.5$ \\
\includegraphics[height=\figwidth]{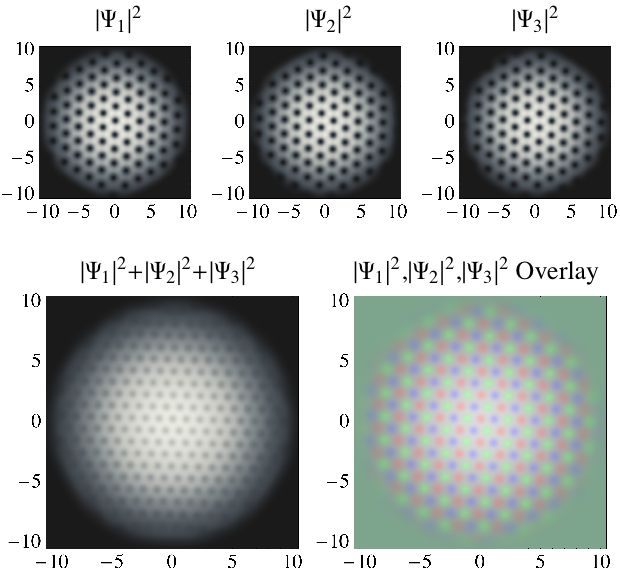} & \includegraphics[height=\figwidth]{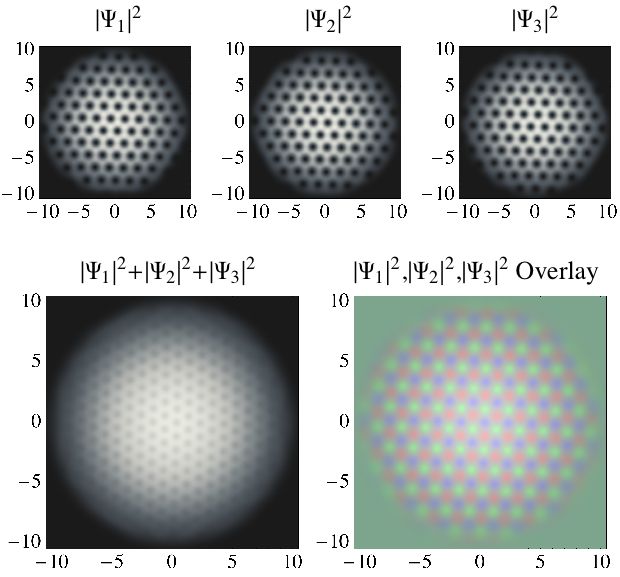} \\ 
	(c) $\delta=0.9$ & (d)  \\
\includegraphics[height=\figwidth]{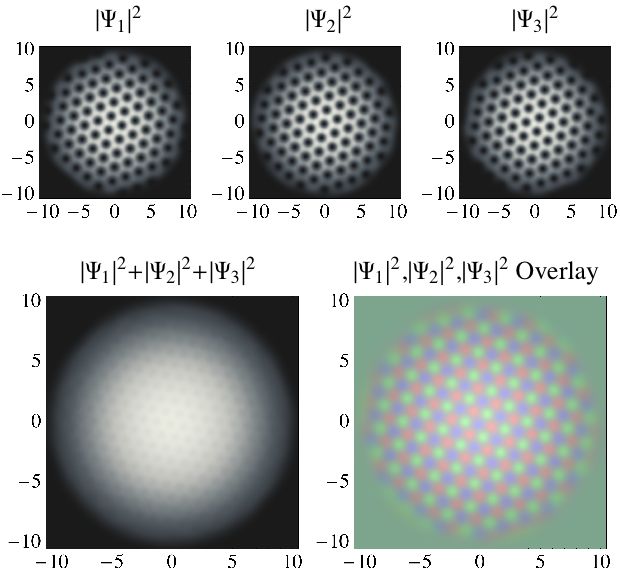} &
\includegraphics[width=\figwidth]{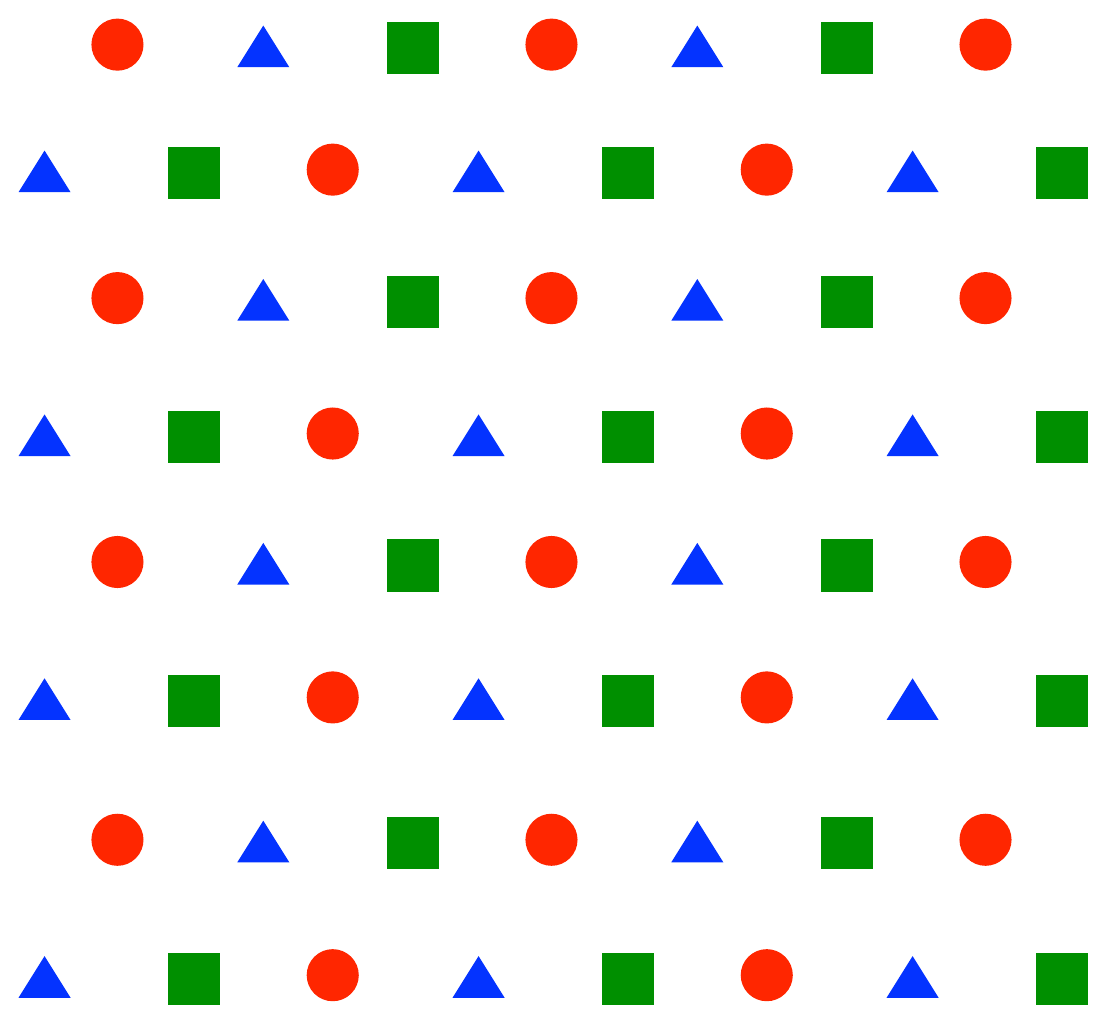}\\
\end{tabular}
\caption{
(Color online) (a)--(c) One-third quantized vortex lattice for $g<g'$.
The top pictures in each panel are plots of the single component density $n_{i}=|\Psi_{i}|^{2}$.
The bottom left picture is the full density profile $n=|\Psi_{1}|^{2} + |\Psi_{2}|^{2} + |\Psi_{3}|^{2}$, while the bottom right picture is the same plot, where the various components are distinguished by color (gray tones).
Red regions correspond to zeros in the density of the first component, 
blue regions stay where the density of the second component vanish, 
while green areas are the zeros of the third component density.
By this plot it can be seen that vortices in different components are well separated for this range of parameters.
(d) A schematic picture of the
1/3 quantized vortex lattice for $g<g'$, where the different shapes and colors distinguish the different components.
}
\label{fig:Fig3}
\end{figure*}
\def\figwidth{8cm}

\begin{figure*}
\centering
\begin{tabular}{c}
\begin{minipage}[b]{0.47\textwidth}
	(a) $\delta=1$ \\
\includegraphics[width=\textwidth]{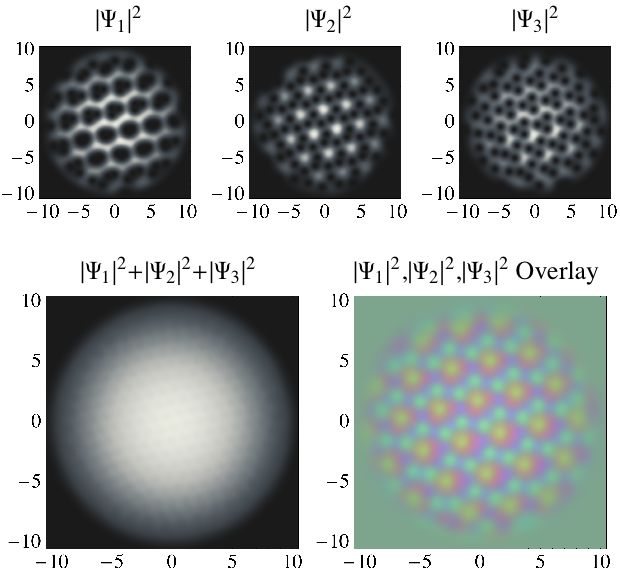} 
\end{minipage}
\hspace{0.5cm}
\begin{minipage}[b]{0.47\textwidth}
	(b) $\delta=1.5$ \\
\includegraphics[width=\textwidth]{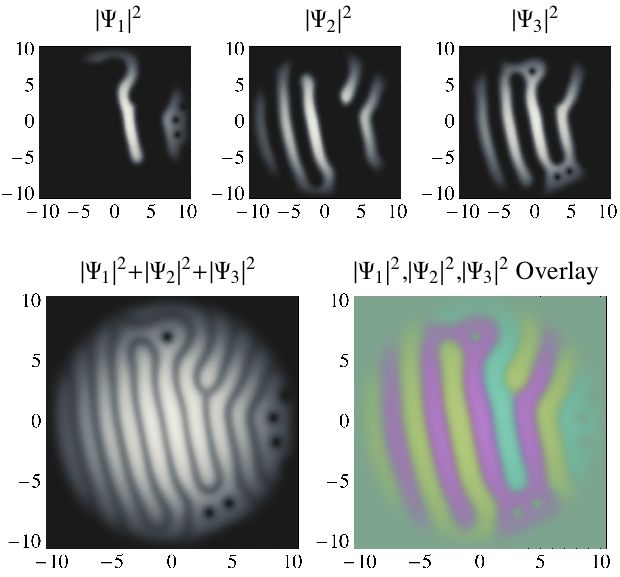} 
\end{minipage}\\
\begin{minipage}[b]{0.47\textwidth}
\begin{tabular}{c}
	\includegraphics[width=\textwidth]{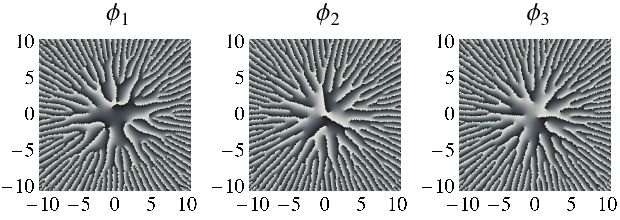} \\ 
	\includegraphics[width=0.7\textwidth]{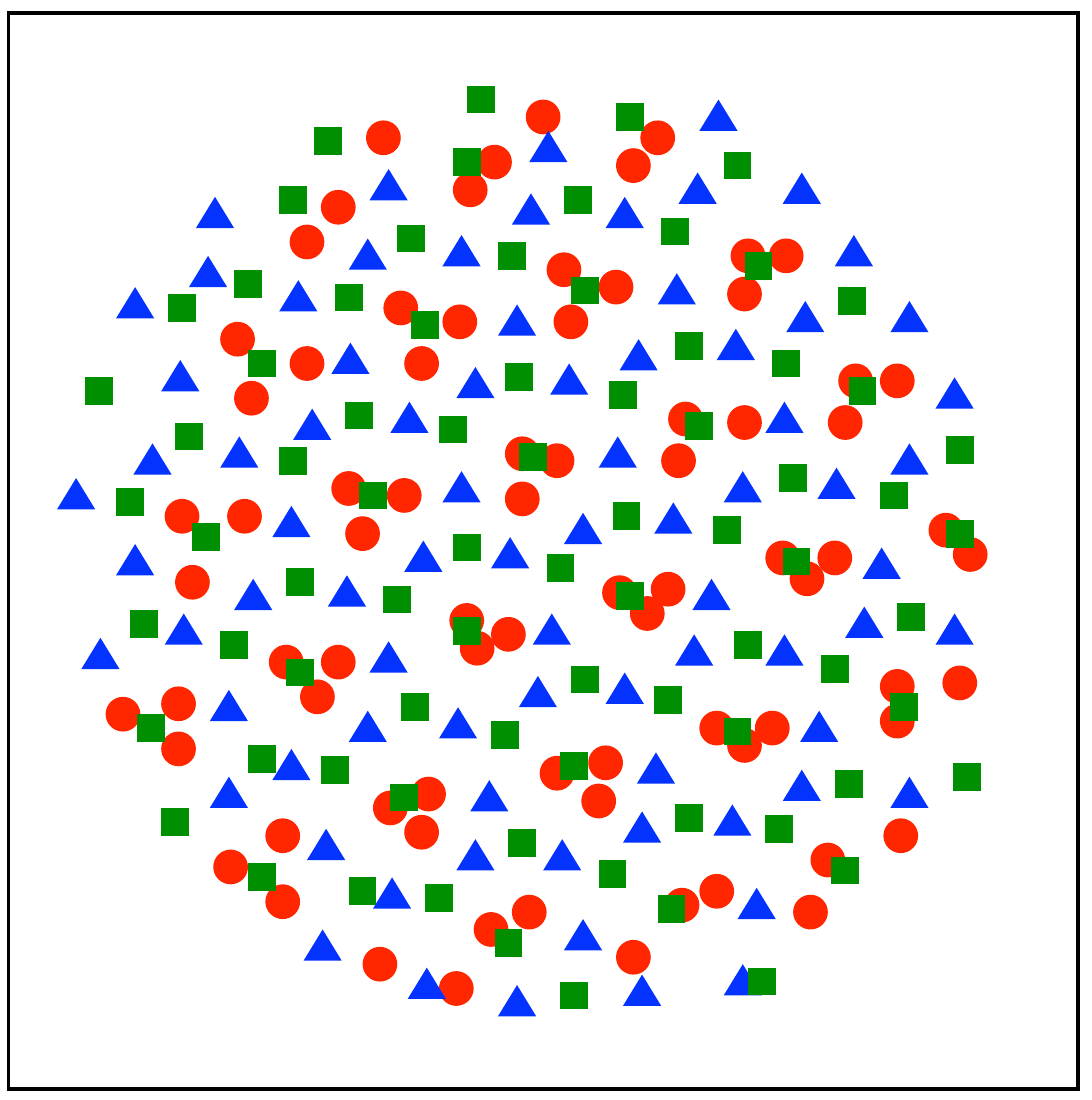}
\end{tabular}
\end{minipage}
\hspace{0.5cm}
\begin{minipage}[b]{0.47\textwidth}
\begin{tabular}{c}
	\includegraphics[width=\textwidth]{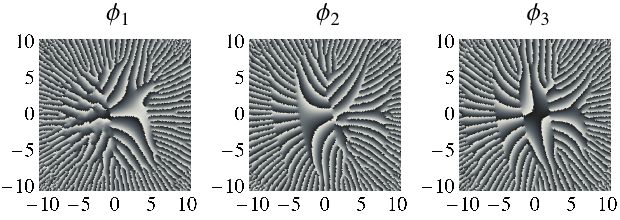} \\ 
	\includegraphics[width=0.7\textwidth]{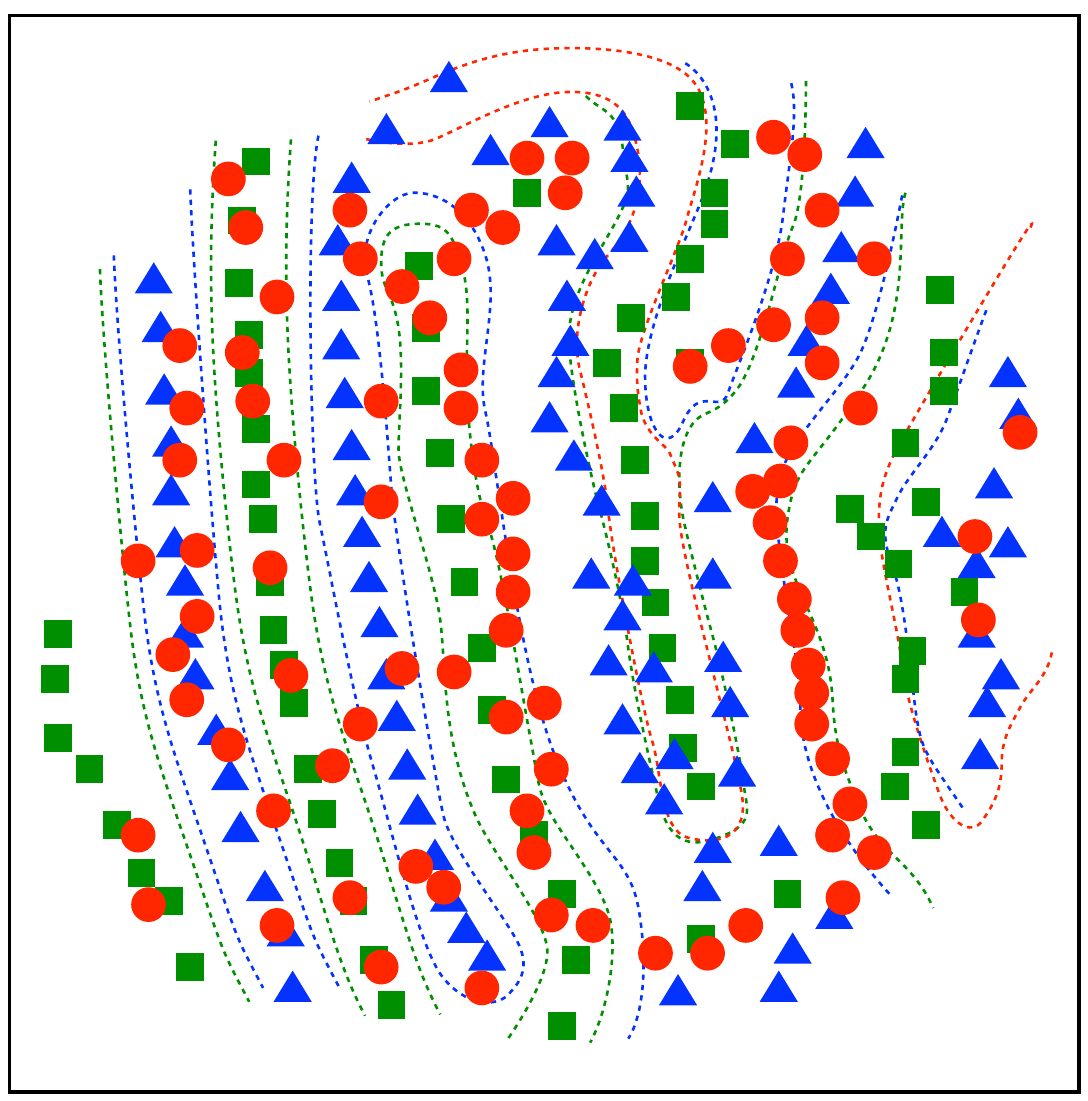}
\end{tabular}
\end{minipage}
\end{tabular}
\caption{
(Color online) 
(a) One-third quantized vortex lattice for $g=g'$ with 
the $U(3)$ symmetry, 
(b) vortex sheets for  $g>g'$. 
From the top in each column we show: plots of the single component density $n_{i}=|\Psi_{i}|^{2}$; plots of the full density profile $n=|\Psi_{1}|^{2} + |\Psi_{2}|^{2} + |\Psi_{3}|^{2}$, where in the rightmost picture the various components are distinguished by color (gray tones), with red corresponding to zeros in the density of the first component, 
blue staying where the density of the second component vanish, 
green corresponding to the zeros of the third component density; a plot of the phase of the order parameter of each component, $\phi_{i}= -i \ln(\Psi_{i}/|\Psi_{i}|)$ 
(black represents zero and white does $2\pi$); a schematic drawing of the positions of vortices distinguished by colors and shapes, with the dashed lines in the right column being the domain walls separating sheets of different components.
Because of the superposition of zeros of the density explained in the text, other colors appear, due to the mixing of the original colors: yellow, cyan and purple correspond to red-green, blue-green and red-blue mixing respectively.
}
\label{fig:Fig4}
\end{figure*}

\section{Gross-Pitaevskii energy functional}
The energy functional of the Gross-Pitaevskii equations for the rotating BECs subject 
to a trapping harmonic potential can be written as:
\begin{align}\label{eq:FreeEnergy}
	E = & \int d^3 r \left\{ \sum_{i=1,2,3} \Psi^*_i \left[ \frac{1}{2} \left( \frac{1}{i} \bm{\nabla} - \Omega \, \hat{\bm{z}} \times \bm r \right)^2 \right. \right. \nonumber \\ 
	& \ \ \ \ \ \ \ \left. + \frac{r^{2}}{2} \left( 1 - \Omega^2 \right) - \mu_i \right] \Psi_i  \nonumber \\
	&  \left. 
+ \frac12 g_1 |\Psi_1|^4 + \frac12 g_2 |\Psi_2|^4 
+ \frac12 g_3 |\Psi_4|^4 \right.  \nonumber\\
&  \left.  + g_{12} |\Psi_1|^2 |\Psi_2|^2 
+ g_{23} |\Psi_2|^2 |\Psi_3|^2 
+ g_{31} |\Psi_3|^2 |\Psi_1|^2 
 \right\} 
\end{align}
where the derivatives include $(r,\theta)$ coordinates.  
We measure distances and energies in terms of $b = \sqrt{\hbar/m\omega}$ and $\hbar \omega$ respectively, where $m$ is the mass of the atoms and $\omega$ is the frequency of the trapping harmonic potential.
As denoted in the introduction, we consider the symmetric case, 
\begin{align}
&& g_{11}=g_{22}=g_{33} \equiv g , 
 \quad 
 g_{12}=g_{23}=g_{31} \equiv g^\prime, \nonumber\\
&& m_1=m_2=m_3 \equiv m, \quad 
    \mu_1=\mu_2=\mu_3 \equiv \mu. 
  \label{eq:couplings}
\end{align}
When $g \neq g'$, the system is $S_3$ symmetric 
under exchanges $\Psi_i \leftrightarrow \Psi_j$.
When $g=g'$, the system is $U(3)$ symmetric 
$\Psi_i \to \Psi_i' = U_i{}^j \Psi_j$ 
with $U \in U(3)$. 

\section{Vortex lattices}
We minimize the free energy~\eqref{eq:FreeEnergy} by 
the non-linear conjugate gradient method 
(the imaginary time propagation) 
in the \textsc{freefem}++ package. 
In all the numerical simulations we take 
$g=1$, $\Omega=0.96$ and $\mu=6$.
We calculate the ground state of the system 
by minimizing the energy functional \eqref{eq:FreeEnergy}.

We first start as the initial configuration from the ground state of a non-rotating BEC 
where no vortices are present.
We fix the rotation speed and let the vortices form in response to the rotation.
We find that the vortices are nucleated as {\it integer} vortices disposed in an Abrikosov lattice
for each value of $g'$. 
This implies that an integer vortex composed of 
three different fractional vortices, 
$\Psi_i \sim e^{i\theta}$ with the other two having no winding, 
is metastable. 
We show some of the results we obtained in Fig.~\ref{fig:Fig1}. 
The defect in the lattice of Fig.~\ref{fig:Fig1} (c) appears by chance and it is metastable. 

Next, we introduce a perturbation in the lattice of integer vortices by splitting the one near the center of the cloud into a set of three different fractional vortices, with very small spacing.
We use this perturbed configuration 
as the initial configuration in the next step and then obtain 
the final configuration for each value of $g'$.
We can expect that in a real situation the fluctuations of the trapped condensate will be responsible for the perturbation needed to eventually break the metastable integer vortices into the fractional ones.
This behavior and the final structure of the three component lattice depend on the sign and on the magnitude of $g'$, the coupling constant of the inter-component interactions. 
To characterize the system, let us define the ratio 
of the inter-component and intra-component couplings by
\begin{align}
	\delta \equiv g'/g.
\end{align}

We examined a wide range of values for $\delta$, from $\delta_{\text{min}}=~-0.3$ to $\delta_{\text{max}}=1.5$, with the separation $\Delta\delta=0.1$.
In Fig.~\ref{fig:Fig2}, we show the results we obtained for $\delta<0$.
We obtained an Abrikosov lattice of integer vortices, as can be seen from the bottom pictures in each panel.
When $\delta<0$, in fact, the interaction between the vortices belonging to different components is attractive 
\cite{Eto:2011wp} 
and, as expected, only integer vortices are formed and they arrange in an Abrikosov triangular lattice.

When $\delta=0$ each component is decoupled from the others and vortices of different components organize in an Abrikosov lattice independently.

The repulsive inter-component interaction present for $0<\delta<1$ \cite{Eto:2011wp} makes energetically convenient for the system to form fractional vortices instead of integer ones.
The results obtained for this range of values for $\delta$ are reported in Figs.~\ref{fig:Fig3}(a)--\ref{fig:Fig3}(c). 
The schematic picture of the whole vortex lattice 
structure is drawn in Fig.~\ref{fig:Fig3}(d). 
As can be seen from these figures, the fractional vortex lattice does not show any defect and both the total and individual lattices have the Abrikosov structure.
This fact states the robustness of the shape of the fractional vortex lattice in a three-component BEC.
One can find that vortices in each component constitute the Abrikosov lattice. 
Moreover, the repulsive force among vortices in different components determines relative positions of 
the Abrikosov lattice of each component, 
and consequently the total configuration is also 
in the form of the the Abrikosov lattice. 
This is possible because the number of components is three. 
In fact, in two-component BECs, there are two kinds of 
the vortex lattice structures depending on the inter-component coupling, that is, 
the Abrikosov's triangular lattice and square lattices.
On the other hand, in our case of three component BEC, 
the Abrikosov lattice structure is robust in all ranges of 
the inter-component couplings for $0<\delta<1$.

If $\delta = 1$, the symmetry of the system is enhanced to 
a $U(3)$ symmetry.  
A result is shown in Fig.~\ref{fig:Fig4}(a). 
The situation is between the ordered Abrikosov lattice for $\delta<1$ 
and the vortex sheets for $\delta > 1$.
A new feature can be seen in this case: 
Density profiles of different components 
can have overlapped (almost) zeros.
The right plot in the second row in Fig.~\ref{fig:Fig4}(a) shows a superposition of the density profiles
of each component, distinguished by red, blue and green colors.
However, other colors are present, which are generated by the mixing of the original ones: Yellow comes from the mixing of red and green, purple is the result of the mixing of red and blue, while cyan arises from the mixing of blue and yellow.
In order to specify the positions of vortices,
we can look at the plot of the phase of each of the order parameters, defined as $\phi_{i}=-i \ln (\Psi_{i}/|\Psi_{i}|)$.
By using a schematic representation, reported in the last row of Fig.~\ref{fig:Fig4}(a), we can compare the positions of the vortices in each component and can see that vortices are not coincident, even if the (almost) zeros of the density are overlapped. 
This happens because two vortices with different components are close to each other in some place, 
although they never coincide.
We also notice some symmetric structure in the lattice.  

When $\delta > 1$, the phase separation occurs and 
vortex sheets appear as in Fig.~\ref{fig:Fig4}(b).
This resembles the cases of two-component BECs, 
but new features arise. 
In this case, in a region where one component is present, 
the other two components are absent 
because of the phase separation 
so that 
zeros of the density profiles of the other two components are completely overlapped, 
as can be seen from the complete mixing of colors in the rightmost picture of the second row of Fig.~\ref{fig:Fig4}(b). By using the phase plot as done for the $\delta=1$ case, 
we observe that in a region where one component is nonzero and 
the amplitudes of the other two components are zero,
there are ghost vortices of these two components    
whose positions are not coincident.

\begin{figure}
	\includegraphics[width=8.5cm]{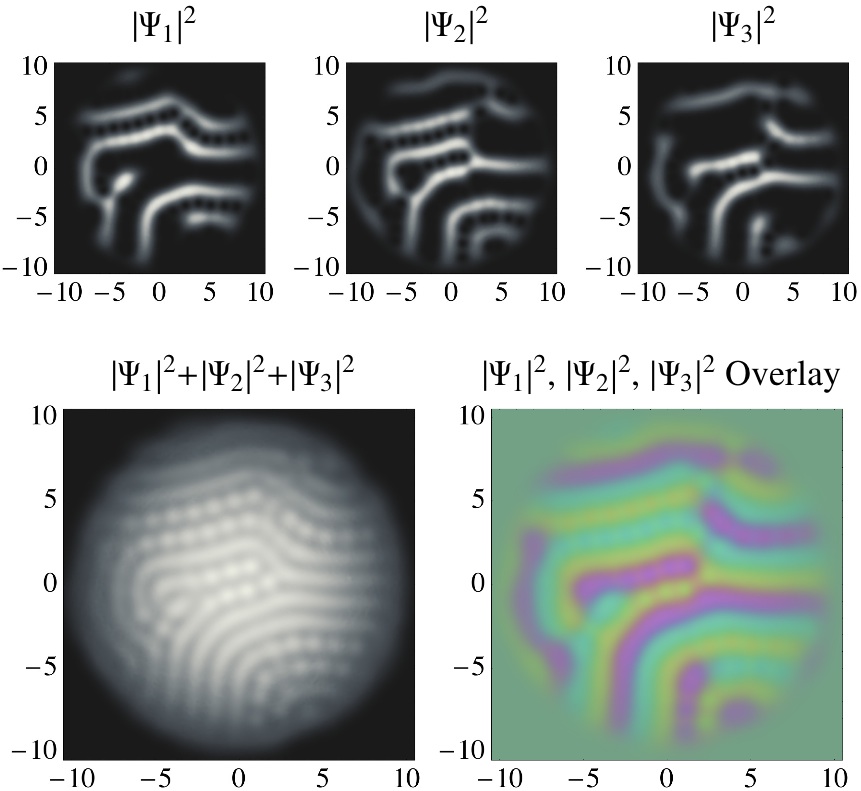}
	\caption{(Color online) The results obtained for $\delta=1.05$. 
	The top pictures are plots of the single component density $n_{i}=|\Psi_{i}|^{2}$.
	The bottom left picture is the full density profile $n=|\Psi_{1}|^{2} + |\Psi_{2}|^{2} + |\Psi_{3}|^{2}$, while the bottom right picture is the same plot, where the various components are distinguished by color (gray tones).
	Vortex sheets are formed for values of $\delta$ just above 1.
	}
	\label{fig:Fig5}
\end{figure}

Moreover, this transition between an ordered lattice structure and the vortex sheets is quite sharp.
In fact we report in Fig.~\ref{fig:Fig5} the results obtained for $\delta$ slightly greater than 1, namely $\delta=1.05$.
As can be seen form the plots, the vortices in the different components are not arranged in an ordered lattice, but they constitute vortex sheets.

\section{Summary and discussion}
In conclusion we have found a robustness 
of the ordered Abrikosov's triangular lattices
of fractional vortices 
in three-component BECs 
in a large parameter region,  
where the inter-component coupling is less 
than the intra-component coupling $g'<g$. 
Our finding is  
in contrast to the cases of two-component BECs, 
where triangular and square lattice are present 
depending on the inter-component coupling 
constant and the rotation speed. 
In the range $g\leq g'$ $(\delta<1)$, 
we find that (almost) zeros of the density
of different components can coincide, even if the vortices lying inside these regions are separated.
When $g=g'$ $(\delta=1)$, two vortices in different components are close in some places, where we see the overlap of almost zeros of these two components.
When $g < g'$ $(\delta >1)$, vortex sheets appear similarly to the case
of two component BECs. 
In a region where one component is present, 
the other two components must vanish because of the phase separation, and in that region there are ghost vortices in these two components whose positions do not coincide.

Our results imply that 
in the region $0<\delta<1$
triangular structures of 
colorful vortex lattices in three components are very robust, 
in contrast to the cases of two components. 
This happens because the number of components, 
three, coincides with the number of vertices of a triangle.
The implication to a colorful vortex lattice in the color superconductor 
is suggestive.  
Since both the number of color and the flavors (up, down, strange) 
 are three, it implies a very robust colorful vortex lattice. 

Because of our motivation to simulate a color superconductor, 
we restrict to the case 
of the symmetric couplings in Eq.~(\ref{eq:couplings}).
Apart from this motivation, we could change couplings asymmetrically.
If we change either $g_i$, $g_{ij}$, $m_i$ or $\mu_i$, 
circulations of vortices are not 1/3 quantized anymore.
When we consider multi-component BECs as 
a mixture of different atoms, this happens in general.
The vortex phase diagram in such a case has been studied 
recently for two-component BECs \cite{Kuopanportti:2012}. 
These remain as a future problem.

If we set $g_{23}=g_{31}=0$, the third component $\Psi_3$ 
decouples from the others, reducing to 
two-component BECs of $\Psi_1$ and $\Psi_2$.
There, we have the phase diagram of the two component BECs
\cite{Mueller:2002,Kasamatsu:2003,Kasamatsu:2005,Woo:2008,
Aftalion:2012,Kuopanportti:2012,Cipriani:2013nya}.
It is an interesting future problem 
how the vortex phase of two-component BECs 
is connected to that of a three-component BEC found in this paper  
by gradually increasing the coupling to $\Psi_3$. 

Another interesting problem is to introduce 
internal coherent couplings between different components 
via Rabi oscillations. 
In the case of two-component BECs, 
two different fractional vortices 
are combined by a sine-Gordon domain wall to become 
a two-vortex molecule or a dimer \cite{Kasamatsu:2004}. 
The effects of this term 
in the vortex phase diagram have been studied recently by the present authors \cite{Cipriani:2013nya}. 
For three-components, the introduction of  
internal coherent couplings results in a three-vortex molecule, 
a trimer \cite{Eto:2012rc}, 
and the same for $N$-component results 
in a vortex $N$-omer \cite{Eto:2013}.

In color superconductors of high density quark matter,
modulated phases, 
called as crystalline color superconductors,  
are proposed \cite{Anglani:2013gfu}. 
Our configurations of vortex sheets at $\delta>1$ 
consist of domain walls, their junctions 
and vortices absorbed into them. 
These configurations may be useful to 
understand how vortices are trapped in the
modulations in crystalline color superconductors.

Before closing the paper, let us make a comment 
on a possible realization in experiments. 
Two-component BECs of different hyperfine states of 
the same atom have been already realized using
the $|1,-1\big>$ and $|2,1\big>$ states \cite{Matthews} 
and the $|2,1\big>$ and  $|2,2\big>$ states \cite{Maddaloni:2000} 
of $^{87}$Rb, respectively. 
Systems with three components can be realized 
in principle 
by using a mixture of the above mentioned states 
of $^{87}$Rb 
via an optical trap \cite{Hamner:2011}, 
and our prediction is testable in laboratory experiments.

\section*{Acknowledgements} 
M.~C. thanks the Department of Physics at Hiyoshi, Keio University, for warm hospitality in the beginning of this project.
M.~N. thanks INFN, Pisa, for partial support and hospitality while this work was done.
Both authors thank Walter Vinci and Kenichi Konishi for useful discussions.
The numerical calculations were performed on the INFN CSN4 cluster located in Pisa.
The work of M.~N.  is supported in part by 
a Grant-in-Aid for Scientific Research (No. 23740198 and 25400268) 
and by the ``Topological Quantum Phenomena'' 
Grant-in-Aid for Scientific Research 
on Innovative Areas (No. 23103515 and 25103720)  
from the Ministry of Education, Culture, Sports, Science and Technology 
(MEXT) of Japan. 


\end{document}